ADVENTURES IN CODE

# Commands as AI Conversations

Diomidis Spinellis, AUEB & TU Delft

*Abstract*—Developers and data scientists often struggle to write command-line inputs, even though graphical interfaces or tools like ChatGPT can assist. The solution? "ai-cli," an open-source system inspired by GitHub Copilot that converts natural language prompts into executable commands for various Linux command-line tools. By tapping into OpenAI's API, which allows interaction through JSON HTTP requests, "ai-cli" transforms user queries into actionable command-line instructions. However, integrating AI assistance across multiple command-line tools, especially in open source settings, can be complex. Historically, operating systems could mediate, but individual tool functionality and the lack of a unified approach have made centralized integration challenging. The "ai-cli" tool, by bridging this gap through dynamic loading and linking with each program's Readline library API, makes command-line interfaces smarter and more user-friendly, opening avenues for further enhancement and cross-platform applicability.

T he impressive advances in generative AI, made possible by large foundation models, are driving all sorts of companies to integrate AI capabilities into their products and services. How can we do the same for the tools we use as developers? What are the possible approaches? What does such integration entail from an engineering perspective?

There's certainly a need for AI help. I often struggle to remember the precise incantation that a command-line tool requires and, judging from highly-voted posts on StackOverflow I regularly encounter, I see I'm not the only one. Graphical user interfaces help here, but in terms of capabilities and efficiency, they're often no match for their command-line siblings. ChatGPT is also very helpful, but drafting prompts with appropriate context and copy-pasting its answers to the command line seems wasteful. What would be ideal would be a GitHub Copilot–like app for command-line interface (CLI) tools. Here we'll see the design and construction such a system, called *ai-cli*. It's available as open source software from GitHub at https://github.com/dspinellis/ai-cli. When installed and run, it allows writing natural language prompts directly on the command line of many (unmodified) command-line tools, such as the Bash shell, diverse SQL front-ends, the *gdb* debugger, and the *bc* calculator. The press of a hotkey will tap into OpenAI's GPT model and insert into the editing buffer the prompt converted into an executable command.

## TALKING WITH AI

How can we convert a natural language prompt into an executable command? Many vendors of large language models offer them both through interactive web sessions, such as ChatGPT, and through an API. Here we'll see OpenAI's API, but others are quite similar. Accessing the API involves making an HTTP request with JSON data specifying the prompt for which the user wants an AI response. To provide conversation context, the prompt can consist of several messages, each corresponding to one of three roles: the "user" role containing the human user's requests, the "assistant" role containing the OpenAI's responses, and a "system" role specifying the overall direction for the conversation, for example "Talk to me as if I'm five years old". Providing the context with each request, means that OpenAI's back-end doesn't need to maintain discussion state, and also offers more flexibility to the API user. Other parameters specify the model to use (e.g. GPT 3.5 or GPT 4) and a sampling "temperature", which controls the randomness of the provided response. API access authorization is handled out-of-band through the HTTP Authorization header and a "bearer" token — a secret key that OpenAI provides to each API's user.
In our case an API request might be as follows.

```
{
  "model": "gpt-3.5-turbo",
  "temperature": 0.7,
  "messages": [
    {"role": "system", "content": "You are an assistant who provides executable commands for the bash command-line interface."},
    {"role": "user", "content": "List files in current directory"},
    {"role": "assistant", "content": "ls"},
    {"role": "user", "content": "How long has the computer been running?"}
}
```

In practice, *ai-cli*'s "system" prompt also contains instructions that disable explanations and provide textual answers as comments. Furthermore, the "messages" array starts with three canned program-specific user-assistant exchanges that provide multi-shot priming to the AI model, continues with previously typed commands to provide context, and finally ends with the actual user prompt.

If no error occurs, the response is quite simple, with the "content" part being what we're after.

```
{
  "id": "chatcmpl-7m1tRm1A8uAUPC174mQsjx4ql2n14",
  "object": "chat.completion",
  "created": 1691681377,
  "model": "gpt-3.5-turbo-0613",
  "choices": [
    {
      "index": 0,
      "message": {
        "role": "assistant",
        "content": "uptime"
      },
      "finish_reason": "stop"
    }
  ],
  "usage": {
    "prompt_tokens": 167,
    "completion_tokens": 1,
    "total_tokens": 168
  }
}
```

(The usage tokens are important because they determine the API's use cost. At the time of writing the OpenAI "GPT 3.5-Turbo" model with four thousand tokens context is priced at $0.0015/1k prompt tokens and $0.002/1k completion tokens. Tokens are common words or parts of less common ones.)

## EXAMINING ALTERNATIVES

Now imagine the task of adding AI help to a large set of command-line tools. The work can be simplified by developing a suitable component. However, this still needs to be integrated with each tool. If the tools were developed by a single company, its managers could plan the task and ask developers to implement it within a (typically overoptimistic) deadline. In the open-source world things are more interesting. One possibility is to wait for each tool maintainer to add such functionality. But this can take a lot of time, because many tool maintainers are volunteers with little available free time. A more attractive alternative is to somehow make the change in a central location, so that all tools will automatically pick it up.

A few decades ago, this location would have been between the operating system and each command-line tool. Operating systems offer a line-editing capability, so extending this with a shortcut key offering AI help would instantly make this facility available to all programs accepting input from a terminal line-by-line. Sadly, operating systems were slow and conservative in enhancing their line editing prowess. For example, the only editing functionality that the Unix or Linux kernel offers is the ability to erase an entire line or the last entered character or word. To address this shortcoming, individual programs developed their own more sophisticated line editing functionalities. Many gradually adopted a library, GNU Readline and its BSD-licensed Editline alternative, for handling command-line input editing. Consequently, extending Readline with AI help seems like a good choice. As Readline does not offer a plugin mechanism, extending it involves modifying its source code. This, however, requires cumbersome coordination with its maintainers and operating system distributors for integrating the changes. Besides, it's not clear that Readline should include such a heavyweight component; its manual page already comments "It's too big" in the "Bugs" section. Instead, an approach we can adopt is to write code that gets loaded when a program gets run and sets up Readline to provide AI help.

## TICKLING THE DRAGON'S TAIL

In the 1940s American scientists developed a suicidally risky method for assessing the critical mass of nuclear material. It involved removing the spacers between two vertically stacked half-spheres of fissionable material, and then using a screwdriver as a wedge to keep them apart.

While a scientist manipulated the screwdriver to bring slowly one part closer to the other, a Geiger counter measured the increasing radiation. The procedure was called "tickling the dragon's tail", and, as you can well imagine, one time it didn't end well.

Externally changing the functionality of a running program is a similarly delicate operation. The program, its runtime libraries, and the operating system maintain tons of state associated with its operation: open handles, statically allocated buffers, signal handlers, memory pools, dynamically loaded libraries, CPU and device registers. All these can be easily messed up when manipulating a program's state causing it to misbehave or crash.

To avoid interference, we need to minimize the possibilities for it. This means programming as close as possible to the operating system and the hardware. For example, a *write* call through the operating system interface will be atomically executed, while a similar operation through an input-output library will be typically buffered in uncontrollable ways. Consequently, the most appropriate programming language for such tasks is C, which has a minimalist and ringfenced runtime environment.

Although C has earned its place in the history books, nowadays it's rarely the best programming language choice. For user-facing applications, it's more appropriate to use development frameworks and their corresponding languages, such as Java, C#, and JavaScript. For performance-critical tasks, well-crafted modern C++ can be just as efficient (on the margin) in terms of memory requirements and execution speed as C, while also being a lot more powerful. For many more other data mangling tasks one can be orders of magnitude more productive by using Unix command-line tools or Python scripts. However, for the requirements of *ai-cli*, C fit like a glove.

The operation of *ai-cli* relies heavily on shared libraries. These are a feature of modern operating systems that allows diverse programs to share functionality, such as a language's runtime support, compression, graphics capabilities, cryptography, or (you guessed it) line editing. These libraries contain compiled native code and are loaded on-demand (dynamically) when a program that uses them gets executed. By being shared among multiple programs they save the disk space that would be required for having their code individually (statically) linked to each program. In a modern Debian GNU/Linux operating system distribution about 600 programs share 160 MB of library code, which would otherwise require 1.9 GB if it was linked individually to each one of them. Some systems even have processes share among them their library code as it is loaded on main memory, resulting in additional RAM savings.

The *ai-cli* code attaches to the editing part of command-line tools through mechanisms associated with shared libraries. The first is the LD_PRELOAD environment variable. Environment variables are key-value pairs maintained by the operating system and stored in the memory image of each running process. They can be set by the shell and are inherited when a parent process executes a child one. They are used for passing arbitrary information down the process tree: things ranging from the path in which commands are searched to the user's preferred editor. The LD_PRELOAD environment variable instructs the code that loads shared libraries to load one or more additional libraries before running each program. It is often used for debugging and instrumentation. In our case this is used to load the *ai-cli* code together with each program.

The second mechanism used for attaching *ai-cli* to external programs is a compiler extension qualifier "__attribute__((constructor))" that can be associated with C and C++ functions. When the dynamic library linker encounters functions compiled with this qualifier in a shared library, it transfers control to them before executing the program's main function. In the case of *ai-cli* its initialization function is marked in this way. When it gets executed, it uses the API of the dynamic loader library to see if the program is linked with Readline and can therefore be extended with AI help functionality. If so, it creates a new Readline function that provides AI help and binds it to corresponding keystrokes ("Ctrl-X A" by default).

## REFINING

The development of *ai-cli* faced some important unknowns. Could OpenAI's API provide plain and appropriate executable commands as suggestions? Could *ai-cli*'s code hook onto another program and manipulate its Readline interface to add AI help? The first question's answer was obtained with a 25-line Python program that talked to OpenAI through Python's *requests* API allowing experimentation with prompts and answers in the way that would be later implemented in C. The second answer came through a 40 line proof-of-concept C program that attached itself to other command-line programs and hooked to their Readline API. This code provided a fixed response ("The answer to … is 42") to lines starting with "ai ". You can find that code in the *ai-cli* repository's first commit. Little remains from that code, but it demonstrated that a generic AI assistant for arbitrary command-line programs was indeed viable.

Many elements of the *ai-cli*'s operation need to be configured at runtime. Therefore, the next step involved adding a flexible and structured configuration file format and the ability to read multiple configuration files. The investment quickly paid back, because configuration ended up covering facets ranging from the employed model and the sampling temperature to per-program multi-shot prompts, keystroke mappings, and the OpenAI API key.

*Ai-cli* is quite small in terms of code size. In addition to the configuration code (249 lines), the other major elements of *ai-cli* are its initialization code for adding key bindings to Readline (123 lines), code that assembles the JSON for the OpenAI API requests and parses the response (218 lines), and some functions that support the safe handling of memory and character strings (282 lines). The main difficulty was not churning out the required code but understanding and addressing various subtle issues.

Dynamically attaching to third-party programs turned out to be trickier than what the proof-of-concept program emonstrated. The linking would fail with unresolved global variable references if the third-party program lacked the libraries *ai-cli* was using, such as *libcurl* used for making the OpenAI HTTP requests. This happened because, although it's easy and efficient to dynamically patch function calls at runtime, it's more cumbersome and inefficient to do this for arbitrary global variable references. Consequently, dynamic linking does not offer this functionality. The problem was addressed by dynamically obtaining the variables' memory addresses through the dynamic linker API. Thankfully, with C it's trivial to access data based on its memory address by using a pointer.

A more perplexing problem was that arbitrary programs, such as the Unix *man* (manual page) command, would crash when loading *ai-cli.* Understanding why the working *ai-cli* code would fail when linked with some specific programs that weren't even using it was challenging. In the end the mystery was solved through system call tracing and post-mortem debugging of the crashed programs' memory images (the so-called core dumps). It turned out that these programs set up a Secure Computing environment (seccomp) to protect themselves against attacks that could be embedded in the untrusted data they often handled. This environment will forcibly terminate a process if it issues an operating system call outside the known small set of calls the process normally makes. The initialization of the *libcurl* library, issued such a system call (*getrandom*), which resulted in the crash. The solution involved dynamically loading the *libcurl* library only when *ai-cli* gets attached to programs featuring command-line editing, which may actually make OpenAI API calls. This reduces the chance of interference, keeping all programs happy.

Programming in C can be tricky, and so some unit tests (228 lines) helped iron out bugs and gain confidence in the code. The last piece of code written for *ai-cli* was the one to call the OpenAI API. With the Python prototype at hand, it was the one with the fewest unknowns. This turned out to be indeed the case. When all *ai-cli* parts were hooked together, *ai-cli* gave a correct response to its first natural language prompt.

In actual use, *ai-cli* responses to prompts appear in about one second and their quality is on par with what we would expect to get from ChatGPT. The overhead of attaching *ai-cli* with each process invocation is about 70 microseconds — negligible for interactive use.

## FORGING AHEAD

The *ai-cli* program can be extended in many ways. As it's an open-source program hosted on GitHub, it's easy to contribute enhancements through pull-requests (Git patches managed through GitHub's web interface). The easiest extension is the addition of more multi-shot prompts that can optimize responses for diverse command-line tools. This can be easily done by adding more sections to its main configuration file. *Ai-cli* currently supports GNU/Linux (tested natively on the x86_64 and armv7l CPU architectures and on the Windows Subsystem for Linux) and macOS's Homebrew ports. Extending it to work on additional distributions and operating system flavors will allow more people to benefit from it. It would also be interesting to support more large language models. The OpenAI interface is abstracted in a single file and can easily live side-by-side with various alternative APIs. An enticing possibility would be to develop one querying a freely-available model, such as Llama-2. This could be hosted on an organization's server, simplifying access for all its users, and opening the possibility of fine-tuning it for specific use cases. We certainly live in exciting times!

**Diomidis Spinellis** is a professor in the Department of Management Science and Technology at the Athens University of Economics and Business and a Professor of Software Analytics in the Department of Software Technology at the Delft University of Technology. He is a senior member of the IEEE. Further information about him can be found at https://www.spinellis.gr. Contact him at dds@aueb.gr.